\documentclass{jaa}

\usepackage{graphicx}
\usepackage{url}

\begin{document}\sloppy

\title{Quasar catalogue for the astrometric calibration of the forthcoming ILMT survey}


\author{Amit Kumar Mandal\textsuperscript{1,4}, Bikram Pradhan\textsuperscript{2}, Jean Surdej\textsuperscript{3}, C. S. Stalin\textsuperscript{4}, Ram Sagar\textsuperscript{4}, Blesson Mathew\textsuperscript{1}}
\affilOne{\textsuperscript{1}Department of Physics, CHRIST (Deemed to be University), Hosur Road, Bangalore 560029, India.\\}
\affilTwo{\textsuperscript{2}Aryabhatta Research Institute of Observational Sciences, Manora Peak, Nainital 263001, India.\\}
\affilThree{\textsuperscript{3}Space science, Technologies and Astrophysics Research (STAR) Institute, Universit\'e de Li\`ege, All\'ee du 6 Ao\^ut 19c, 4000 Li\`ege, Belgium.\\}
\affilFour{\textsuperscript{4}Indian Institute of Astrophysics, Block II, Koramangala, Bangalore, 560034, India.}

\twocolumn[{

\maketitle

\corres{amitkumar@iiap.res.in}

\msinfo{27 May 2020}{13 July 2020}

\begin{abstract}
Quasars are ideal targets to use for astrometric calibration of large scale 
astronomical surveys as they have negligible proper motion and parallax. 
The forthcoming 4-m International Liquid Mirror Telescope (ILMT) will survey the sky that covers a width of about 27'. To carry out astrometric 
calibration of the ILMT observations, we aimed to compile a list of quasars 
with accurate equatorial coordinates and falling in the ILMT stripe. Towards 
this, we cross-correlated all the quasars that are known till the present date 
with the sources in the {\it Gaia}-DR2 catalogue, as the {\it Gaia}-DR2 sources have 
position uncertainties as small as a few milli arcsec ($mas$). We present here the 
results of this cross-correlation which is a catalogue of 6738 quasars that 
is suitable for astrometric calibration of the ILMT fields. In this work, we 
present this quasar catalogue. This catalogue of quasars can also be used to 
study quasar variability over diverse time scales when the ILMT starts its 
observations. While preparing this catalogue, we also confirmed that  quasars 
in the ILMT stripe have proper motion and parallax lesser than 20 $mas ~yr^{-1}$ and 10 $mas$, 
respectively. 

\end{abstract}


}]

\doinum{12.3456/s78910-011-012-3}
\artcitid{\#\#\#\#}
\volnum{000}
\year{0000}
\pgrange{1--}
\setcounter{page}{1}
\lp{1}

\section{Introduction}
The 4 m International Liquid Mirror Telescope (ILMT) which will observe in the 
Time Delayed Integration (TDI) mode is expected to be commissioned soon on the 
Aryabhatta Research Institute of Observational Sciences (ARIES) site in Devasthal, 
India (Surdej et al. 2018). The ILMT will be repeatedly scanning the sky within 
a narrow stripe of width $\sim$27'. The positions of the celestial 
objects in the sky change with time as the sky moves across the fixed detector surface 
due to the rotation of the Earth around the polar axis. The TDI mode of the 
charge couple device (CCD) mounted at the ILMT helps to track the stars by 
registering the electronic charges with the rate at which the target source 
drifts across the detector (Surdej et al. 2018) and the positions of each object 
in the observed image come out in pixel units. To convert the observations from 
the pixel coordinate system to the world coordinate system ($\alpha$, $\delta$) 
we need to carry out astrometric calibration of the ILMT fields. For that, we
choose quasars as astrometric standards because of their negligible proper 
motions (PM) and trigonometric parallaxes.

The number of quasars we know as of today has significantly increased since their 
first identification about six decades ago (Schmidt 1963). This increase is 
mainly due to large surveys such as the Two degree Field (2dF) QSO survey 
(Croom et al. 2004) in the southern sky and the Sloan Digital Sky Survey (SDSS) 
in the northern sky (Abolfathi et al. 2018, P\^aris et al. 2018). Quasars known 
from different surveys are also gathered and put together in the form of 
catalogues, through several releases by V\'eron-Cetty \& V\'eron (2006, 2010) 
and the Million Quasars (Milliquas) catalogue by Flesch (2017). These catalogues 
have quasars from different origins with different accuracies in their optical 
positions. But these catalogues do not provide the errors associated with the 
equatorial coordinate positions as well as other information such as parallax 
and PM. Therefore, quasars taken from these catalogues cannot be 
directly used as sources to carry out astrometric calibration. 

The main motivation of the present work is therefore to construct a catalogue  of quasars
that will be used to carry out astrometric calibration of the ILMT fields by 
including (i) more precise positions of the sources with uncertainties, (ii) 
additional important information such as parallax and PM and, (iii) the photometry of the objects. We describe the data used in this work in 
Section 2. The procedures followed to make the quasar catalogue and the outcome is presented in 
Section 3. Applications of the catalogue are discussed in Section 4 followed by the Summary in the final Section. 

\begin{figure}
\includegraphics[scale=0.7]{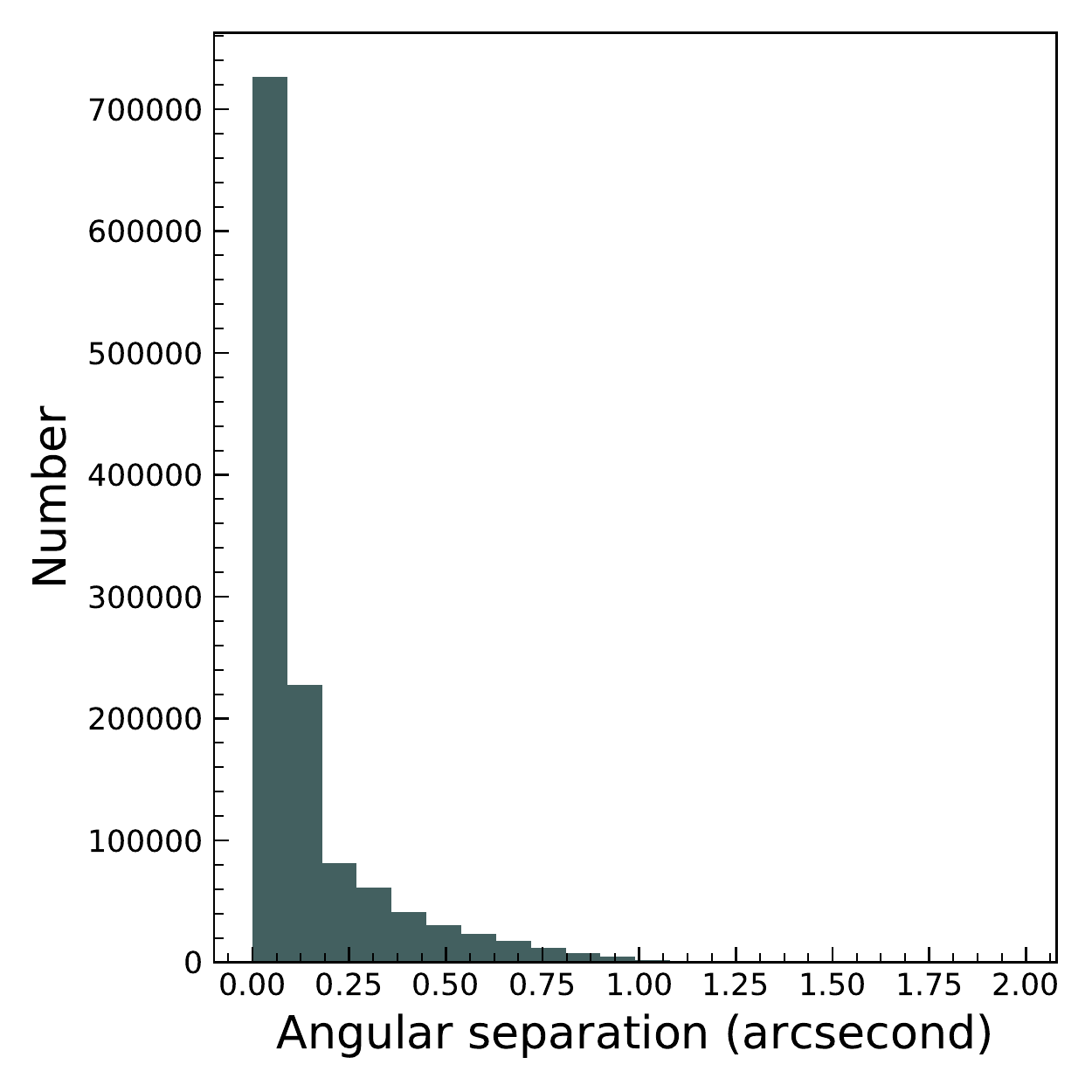}
\caption{Distribution of the angular separation between the quasars in the Million 
Quasars catalogue and their counterparts in {\it Gaia}-DR2.}
\label{fig:fig-1}
\end{figure}

\section{Data used}

To select a catalogue of quasars suitable for astrometric calibration of the ILMT observations
we need to consider all quasars we know as of today. This can come from a wide variety of 
surveys carried out at different wavelengths such as optical, infrared, radio, etc. One such quasar
catalogue suited for our purpose is the Milliquas catalogue. This is the  largest compilation of quasars we have as of today. 
This catalogue contains about 1998464 quasars taken from all the quasar 
surveys available in the literature. The majority of quasars in the Milliquas catalogue comes from the SDSS, one of the ambitious sky surveys covering more than a quarter 
of the sky in the northern hemisphere in five optical filters. Other quasars
included in Milliquas are from the NBCKDE, NBCKDE-v3 (Richards et al. 2009, 2015), 
XDQSO (Bovy et al. 2011; Myers et al. 2015), AllWISE and Peters photometric 
quasar catalogues (Peters et al. 2015) as well as quasars from all-sky 
radio/X-ray surveys. As the Milliquas catalogue is a compilation of various 
quasar surveys, it has varied uncertainties in the equatorial 
coordinates. For astrometric calibration, one needs to have quasars with 
precise positions. A source that provides precise positions of celestial 
sources is the survey being presently carried out by the European Space Agency 
{\it Gaia} mission. {\it Gaia}-DR2 (Lindegren et al. 2018, Gaia Collaboration et al. 2018) 
contains data from the all sky astrometric and photometric survey conducted by 
{\it Gaia} and provides accurate positions for about 1.7 billion sources, 
with PM and parallax measurements for about 1.3 billion sources 
(Marrese et al. 2019). Therefore, to get accurate positions for the quasars 
in the Milliquas catalogue we used the precise and homogeneous measurements from {\it Gaia}-DR2.

\begin{table*}
\caption{The ILMT Quasar (IQ) catalogue. The details on each column of this table is given in Table 2. The full catalogue is available in the electronic version of the article.}
\label{tab:table-1}
\small
\setlength{\tabcolsep}{4pt}
\begin{tabular}{lccccccccccl} \hline
ID-1 & RA & RA-ERR &  DEC & DEC-ERR & z & ... & ... & PM-DEC & PM-DECERR & EPSILON & D \\
(1)* & (2)* & (3)* & (4)* & (5)* & (6) & ... & ... & (16)* & (17)* & (18) & (19)
\\ \hline

2.85518e+18 & 0.06120 & 0.54785 & 29.23513 & 0.28529 & 1.90 & ... & ... & 1.31259 & 0.50398 & 0.00 & 0.00 \\
2.85525e+18 & 0.07183 & 0.22781 & 29.50171 & 0.15339 & 1.40 & ... & ... & 0.27730 & 0.26071 & 0.00 & 0.00 \\
2.85525e+18 & 0.10806 & 0.62917 & 29.50235 & 0.56513 & 2.51 & ... & ... & -1.02912 & 1.06549 & 1.44 & 0.87 \\

\hline
\multicolumn{12}{l}{\footnotesize{ *The values are up to 5 decimal places, the original values retrieved from the {\it Gaia}-DR2 catalogue are mentioned in the IQ catalogue}} \\
\multicolumn{12}{l}{\footnotesize{ available in the electronic version of the article.}}\\
\end{tabular}

\end{table*}

\begin{table*}
\caption{Column information of the ILMT Quasar (IQ) catalogue.}
\label{tab:table-2}
\small
\setlength{\tabcolsep}{3pt}
\begin{tabular}{llccl} \hline
Number & Column Name  & Format & Unit & Description
\\ \hline
1 & ID-1 & String & & Object name as given in {\it Gaia}-DR2
 \\
2 & RA & Double & degree & Right Ascension (J2000) \\
3 & RA-ERR & Double & $mas$ & Error in Right Ascension retrieved from {\it Gaia}-DR2  \\ 
4 & DEC & Double & degree & Declination (J2000) \\ 
5 & DEC-ERR & Double & $mas$ & Error in Declination retrieved from {\it Gaia}-DR2 \\
6 & z & Double & & Redshift \\
7 & ID-2 & String & & Object ID in the Milliquas catalogue \\
8 & TYPE & String & & Classification of the object \\
9 & PROB & Double & & Probability  that the object is a quasar$^\bullet$  \\
10 & MAG & Double & & {\it Gaia} G-band magnitude \\
11 & MAG-ERR & Double & & Error in {\it Gaia} G-band magnitude \\
12 & PLX & Double & $mas$ & Parallax \\
13 & PLX-ERR & Double & $mas$ & Error in parallax \\
14 & PM-RA & Double & $mas \, yr^{-1}$ & Proper motion in RA \\
15 & PM-RAERR & Double & $mas \, yr^{-1}$ & Error in proper motion in RA \\
16 & PM-DEC & Double & $mas \, yr^{-1}$ & Proper motion in DEC \\
17 & PM-DECERR & Double & $mas \, yr^{-1}$ & Error in proper motion in DEC \\
18 & EPSILON & Double & & Astrometric excess noise \\
19 & D & Double & & Significance of excess noise \\

\hline

\multicolumn{5}{l}{\footnotesize{ $^\bullet$ The details on how the probability is assigned to each quasar can be found in Flesch (2015).}} \\

\end{tabular}

\end{table*}

\section{Methods followed and the resulting catalogue}

It is known that quasars represent quasi-ideal astrometric reference sources 
over the celestial sphere because of their negligible PM and are thus suitable 
candidates for use to carry out astrometric calibration (Souchay et al. 2015) of 
the ILMT survey. We therefore aim to gather accurate position, PM and trigonometric 
parallax for all quasars available in the Milliquas catalogue from the 
{\it Gaia}-DR2 database and then select a sub-set of them for the ILMT use.  To 
calculate the absolute or resultant PM $\mu$ we used the relation given by 
Varshni (1982)

\begin{gather}
	\mu = (\mu_{\alpha}^2cos^2\delta + \mu_{\delta}^2)^{1/2}
	\label{eq:pm_eq}
	\end{gather}

where $\alpha$ and $\delta$
are the right ascension (RA) and declination (DEC) respectively. We collected 
$\mu_{\alpha}cos\delta$ and $\mu_{\delta}$ values from the {\it Gaia}-DR2 
database. The error in $\mu$ was calculated using the standard error propagation 
method. To arrive at a separate list of quasars for the ILMT field of view (FoV), 
we followed the following steps:

\begin{enumerate}
\item We cross-correlated nearly 2 million objects in the Milliquas 
catalogue with {\it Gaia}-DR2 with angular proximity of less than 2". We used a 2" angular separation because a large fraction of the objects 
in the Milliquas catalogue are from SDSS that has imaging data with seeing 
less than 2" (Ross et al. 2011). By cross-correlating the Milliquas 
catalogue with the {\it Gaia}-DR2, we arrived at a sample of 1235600 objects 
spanning a range of redshifts up to $z$ = 6.4. The distribution of the angular 
separation for the matched objects between the position in the Milliquas 
catalogue and the position in {\it Gaia}-DR2 is shown in Fig. 1. 
The distribution has a range between 0  and 1.97" with a mean of 0.15"  
and a standard deviation of 0.18". About $99.8\%$ of the objects are 
matched within 1". The distributions of $z$ and G-band brightness of these 
objects are shown in Fig. 2.

\item Among the many parameters provided by {\it Gaia}-DR2, two 
parameters that are relevant for quasar target selection are the 
astrometric excess noise ($\epsilon_i$) and its significance namely the  
astrometric excess noise significance (D). Excess noise $\epsilon_i$ quantifies the 
disagreement between the observations and the best-fitting standard astrometric 
model adopted by {\it Gaia} (Lindegren et al. 2012). A value $\epsilon_i$ = 0 
implies that the source is astrometrically well behaved, and a value $\epsilon_i >$ 0 
indicates that the residuals are statistically larger than expected. However, 
practically there is the possibility that some sources may not behave exactly 
according to the adopted astrometric model. Therefore, the significance of 
$\epsilon_i$ is characterised by its significance namely D (Lindegren et al. 2012). 
If, D $\leq 2$, $\epsilon_i$ is probably not significant and the source may be 
astrometrically well-behaved even if $\epsilon_i$ is 
large\footnote{https://gea.esac.esa.int/archive/documentation/GDR2/Gaia$\_$archive/\\chap$\_$datamodel/sec$\_$dm$\_$main$\_$tables/ssec$\_$dm$\_$gaia$\_$source.html}. Therefore, we only selected sources with D $\leq$ 2 from {\it Gaia}-DR2.
This yielded a total of 1047747 quasars covering the whole sky. For these quasars, 
we calculated PM using Equation 1. The distribution of their PM is shown in 
Fig. 3. From this figure, it is evident that except for a few objects 
(about $0.25\%$) most of them have PM less than 20 $mas ~yr^{-1}$ with a mean 
value and standard deviation of $1.808 \, mas \, yr^{-1}$ and $2.878 \, mas \, yr^{-1}$, respectively.   

\item From the list of quasars obtained at step 2 above, we  made a separate catalogue of 
quasars for the ILMT stripe. The Devasthal observatory where the ILMT is being 
installed is located at a latitude near $29^{\circ} 22' 26"$ (Surdej et al. 2018). 
The width of the ILMT FoV is $\sim$27'. However, the ILMT sky at zenith 
will change with time due to precession as shown in Fig. 4. It has 
been found that if we take a $\sim34'$ wide stripe instead of $\sim27'$, the 
effect of precession during the next 10 years will be taken into account. So as {\it Gaia} 
has a limiting G-band magnitude of 21\footnote{https://www.cosmos.esa.int/web/gaia/dr2}, 
we selected only those quasars having a declination ($\delta$) in the range 
$29.09^{\circ} \leq \delta \leq 29.66^{\circ}$ and G-mag $\leq$ 21 from the 
sample of 1047747 quasars obtained from step 2 since only these will be 
accessible for observations with the ILMT.  Using the above criteria, we arrived
at a sample of 6904 quasars. For slightly less than $2\%$ of these, we do not have the redshift information 
in the  Milliquas catalogue. Excluding those, we arrived at a final catalogue of 6755 
quasars available within the ILMT stripe.

\item A plot of the PM of these objects as a 
function of their G-band brightness  is shown in Fig. 5. From this figure, it
is evident that the majority of the quasars have PM $<$ 20 $mas \, yr^{-1}$. We only found 17 quasars in this list  with PM $>$ 20 $mas \, yr^{-1}$ and all of them are 
fainter than 19.5 mag in the G-band. The nature of these 17 objects could not be 
ascertained due to the lack of optical spectra for them.  Therefore, we neglected those
17 quasars from our list and arrived at a final sample of 6738 quasars that will 
be visible with the ILMT and could be used as astrometric calibrators.

Varshni (1982) 
claimed the existence of high PM quasars namely PHL 1033, LB 8956 and LB 8991 with
PM values of 0.049 $\pm$ 0.013, 0.061 $\pm$ 0.018 and 0.050 $\pm$ 0.018 $arcsec \, yr^{-1}$
respectively. We checked the PM of these objects in {\it Gaia}-DR2 and found
PM values of 0.121 $\pm$ 0.435, 0.188 $\pm$ 0.151 and 0.056 $\pm$ 0.072 
$mas ~yr^{-1}$ for PHL 1033, LB 8956 and LB 8991 respectively. This along with
the observations in  Fig. 5 point to quasars having 
PM $<$ 20 $mas \, yr^{-1}$.  

\item The distributions of redshifts, G-band magnitude and parallax of the ILMT quasars
are illustrated in Fig. 6. They span redshifts up to $z$ = 4.9. Their distribution 
in the galactic coordinate system is shown in Fig. 7. The sample catalogue and the description of its columns are given in Table 1 and Table 2, respectively. The full catalogue is available in the electronic version of the present article.
\end{enumerate}

\begin{figure}[!ht]
\begin{center}
\includegraphics[scale=0.65]{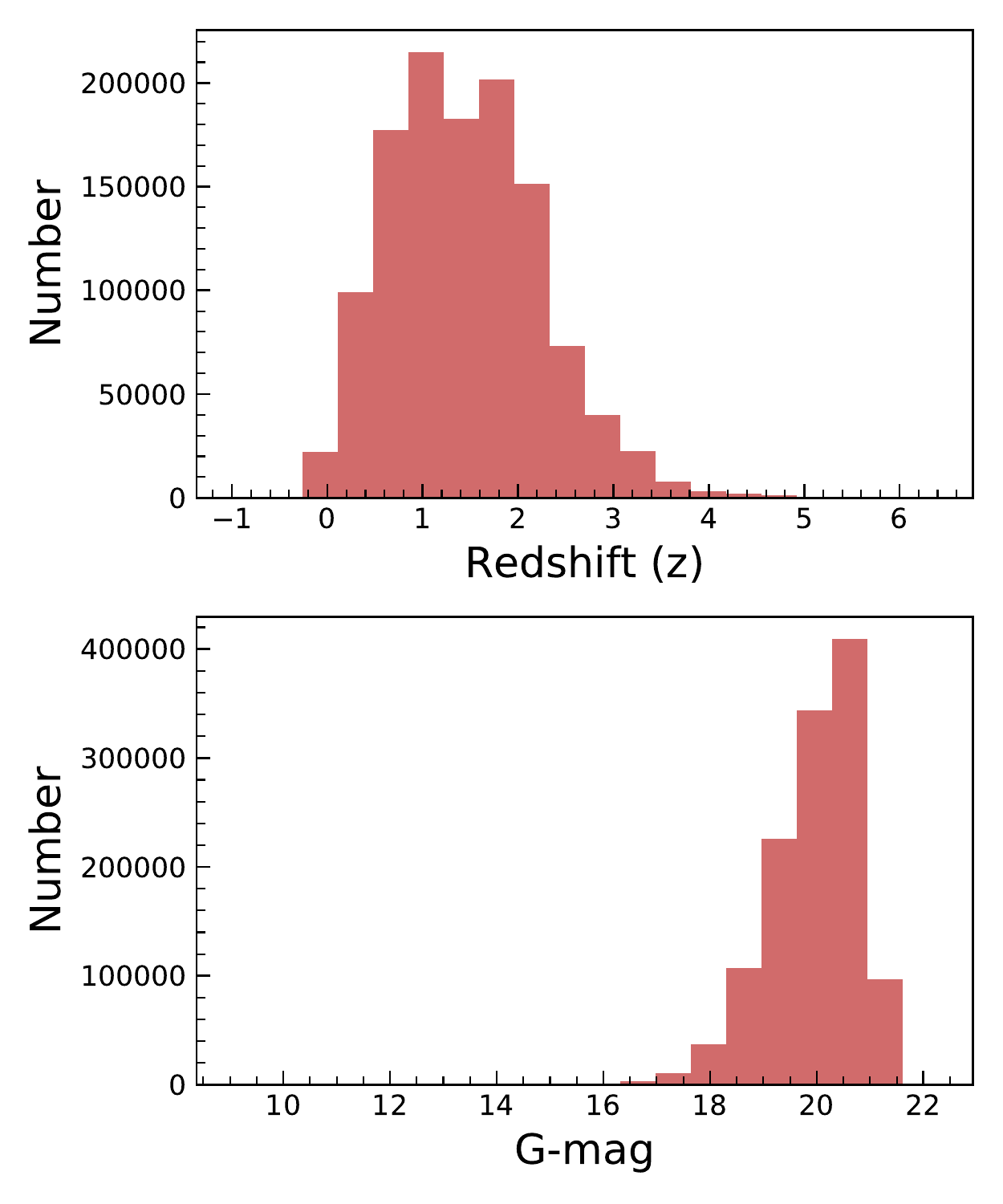}
\end{center}
\caption{Distribution of redshift (top) and {\it Gaia} G-band magnitude (bottom) 
of the quasars selected from the Million Quasars catalogue.}
\label{fig:fig-2}
\end{figure}

\begin{figure}[!ht]
\includegraphics[scale=0.65]{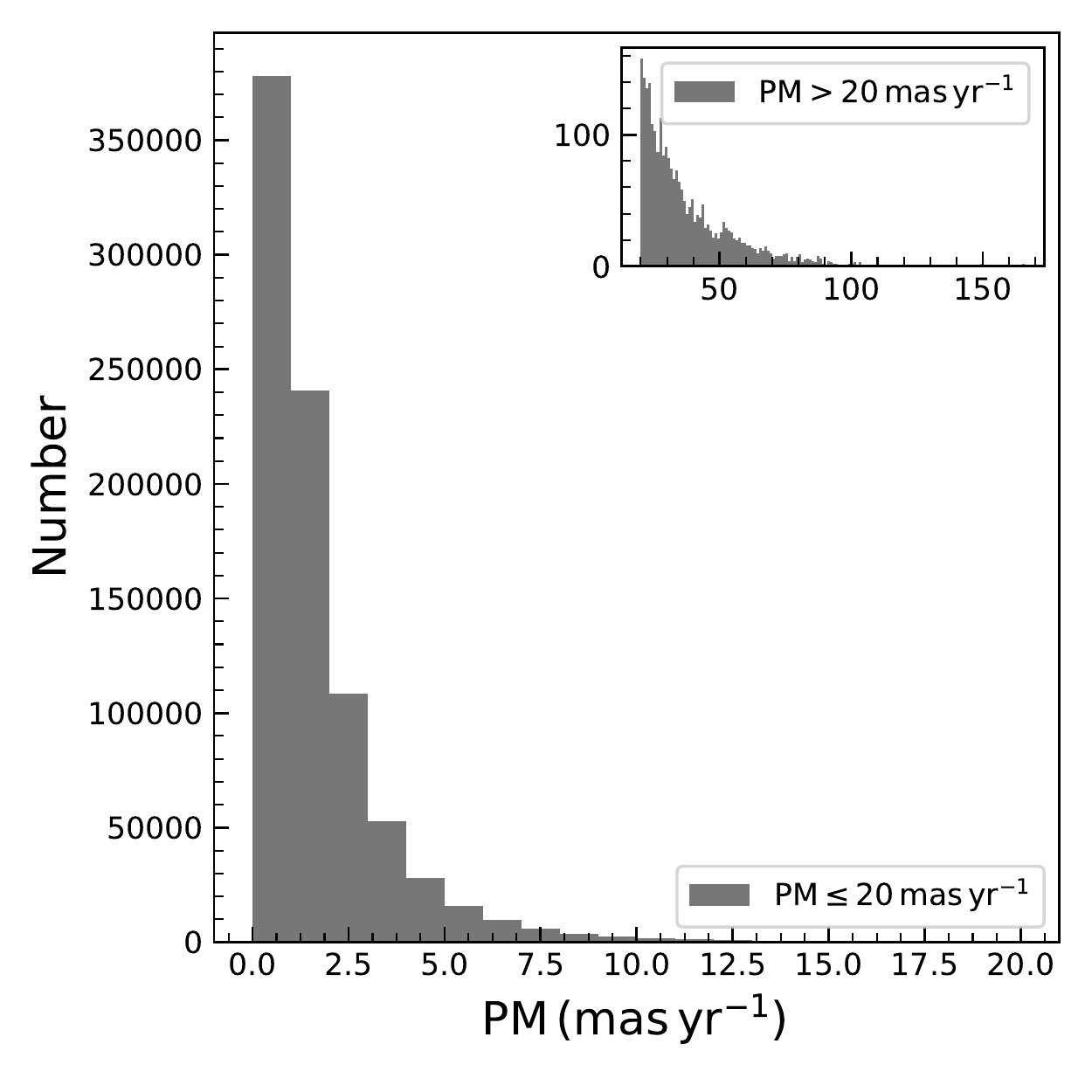}
\caption{Distribution of the PM for quasars in the Million Quasars catalogue with D $\leq$ 2. There are about 2615 quasars with PM $>$ 20 $mas \, yr^{-1}$, their PM distribution is shown in the small box on the same figure.}
\label{fig:fig-3}
\end{figure}

\begin{figure}[!ht]
\includegraphics[scale=0.5]{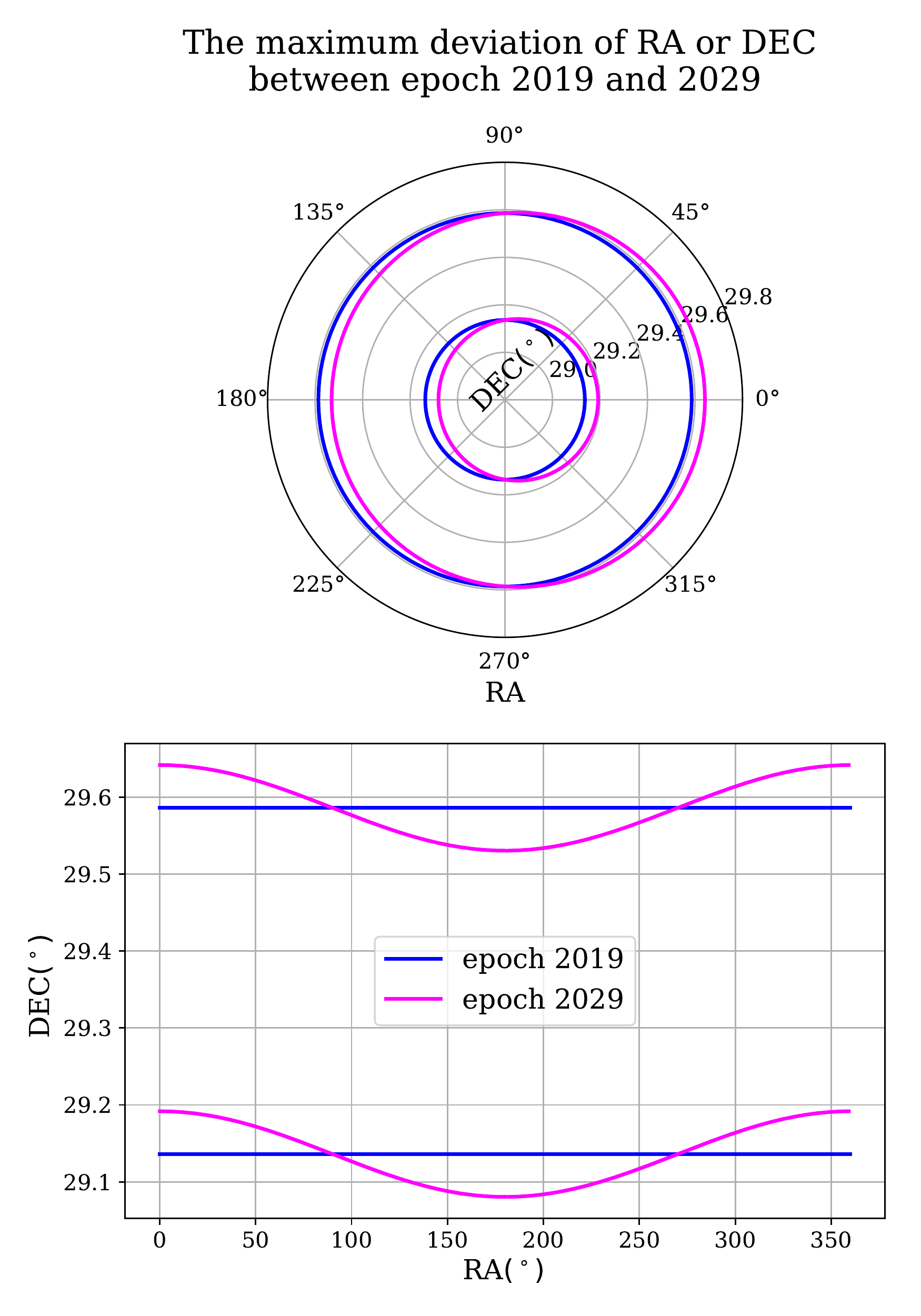}
\caption{(Top) Polar plot showing the astrometric deviation of the ILMT stripe between the 2019 
and 2029 epochs due to precession. (Bottom) Deviation in astrometry between the 
2019 and 2029 epochs considering that the original ILMT stripe is rectangular in 2019 (blue lines) and pink lines represent the same in 2029.}
\label{fig:fig-4}
\end{figure}

\begin{figure}[!ht]
\hspace*{-0.3cm}\includegraphics[scale=0.6]{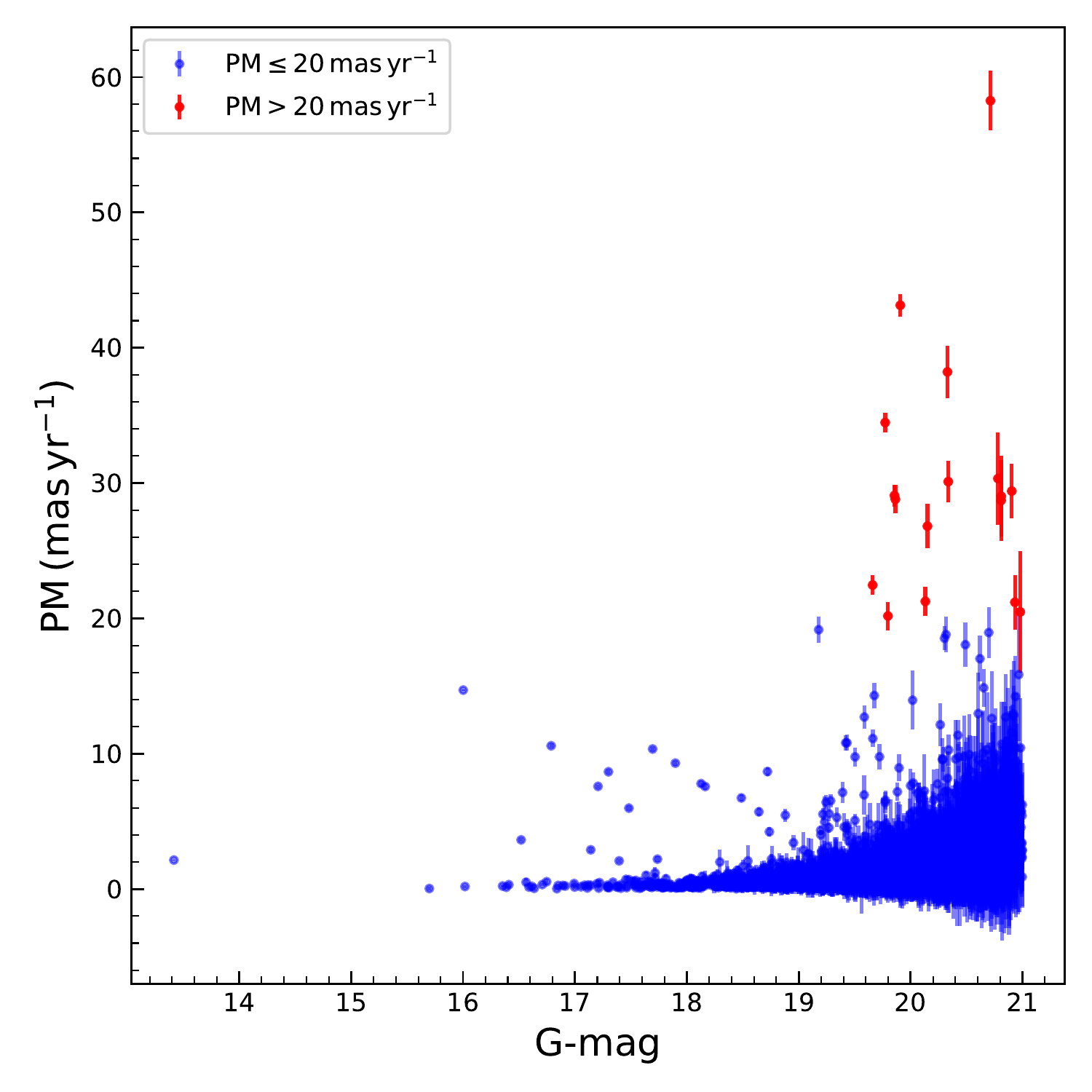}
\caption{Proper motion versus G-mag of the quasars in the ILMT stripe.}
\label{fig:fig-5}
\end{figure}

\begin{figure}[!ht]
\includegraphics[scale=0.8]{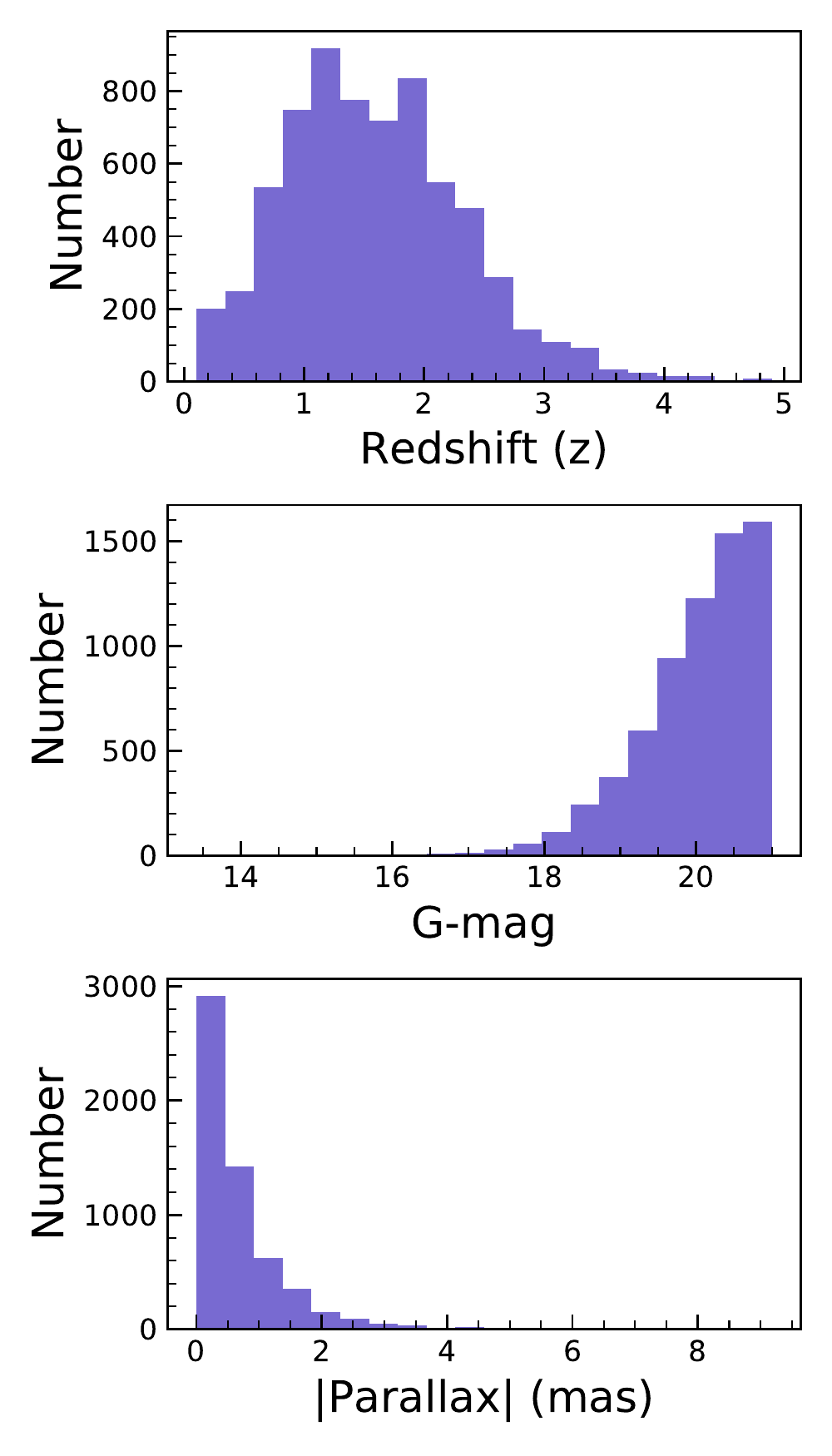}
\caption{Distributions of the redshift, G-band magnitude and absolute parallax of the selected quasars in the ILMT stripe.}
\label{fig:fig-6}
\end{figure}

\begin{figure}[!ht]
\includegraphics[scale=0.52]{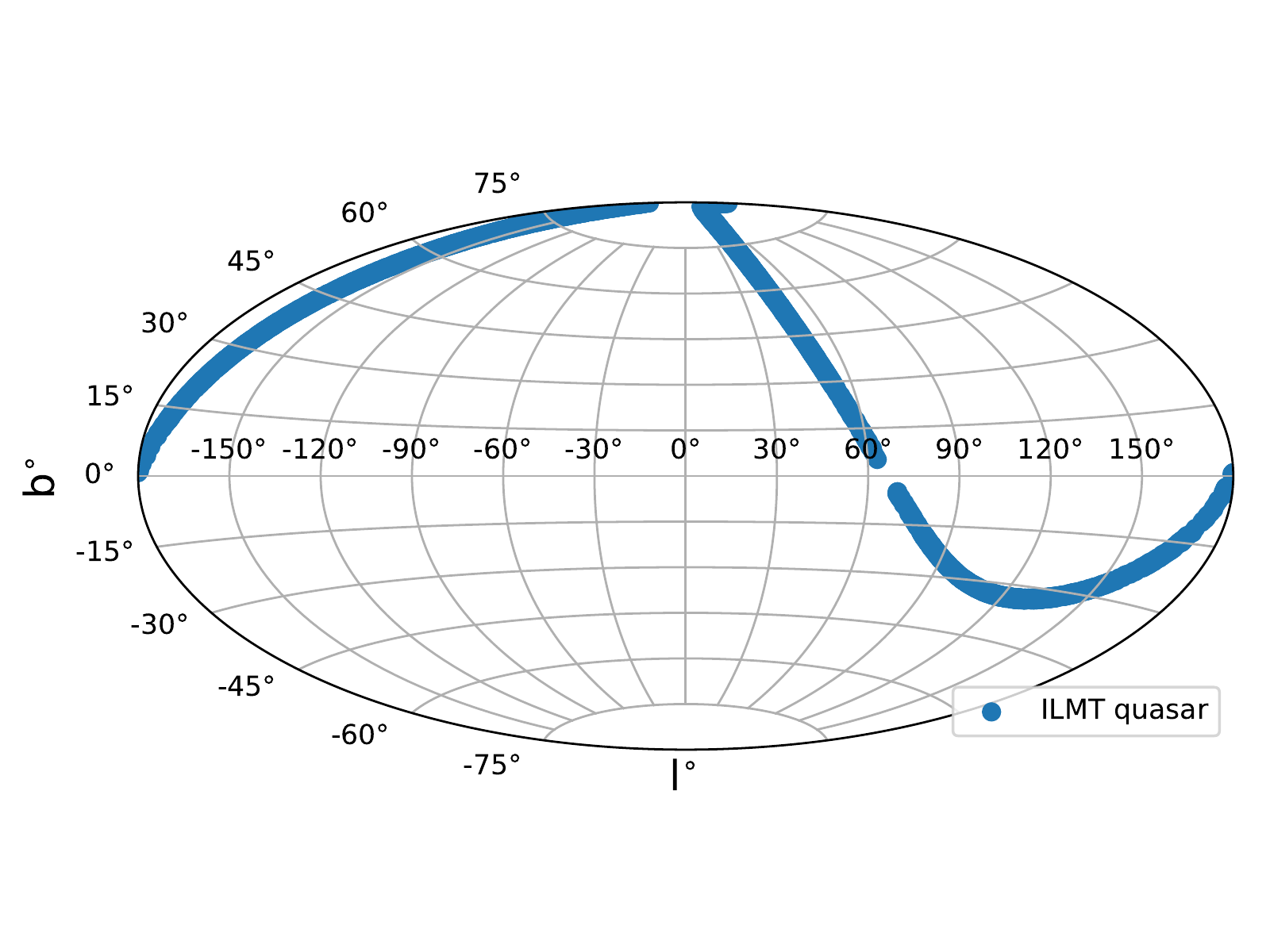}
\caption{Sky distribution of the selected ILMT quasars in the galactic coordinate system. The real surface density of quasars is not considered in this plot.}
\label{fig:fig-7}
\end{figure}

\section{Applications of the catalogue}

The ILMT will be continuously scanning the sky passing over zenith. Such observations will be of interesting use for a wide range of astrophysical applications such as the detection of many extragalactic objects like supernovae, galaxy clusters, active galactic nuclei (AGN)/quasars, gravitationally lensed systems  etc.
(Surdej et al. 2018). Also, as zenith region of the sky will be repeatedly scanned by the ILMT, accumulated observations will be very useful for photometric variability studies of different types of celestial sources. As we have already
arrived at a catalogue of 6738 quasars that will be covered by the ILMT, we describe
below some of the potential applications of this quasar catalogue. 
 
\begin{figure}[!ht]
\includegraphics[width=0.5\textwidth]{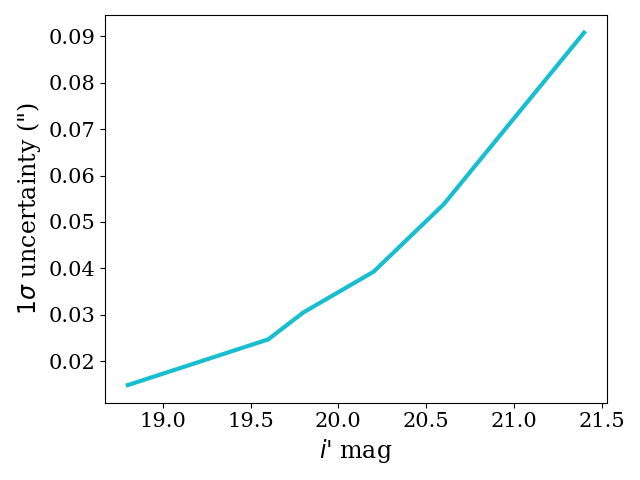}
\caption{Estimation of the $1\sigma$ uncertainty in the astrometric position of point sources with different magnitudes in the ILMT CCD images.}
	\label{fig:fig-8}
\end{figure}

\begin{figure}[!ht]
\includegraphics[width=0.5\textwidth]{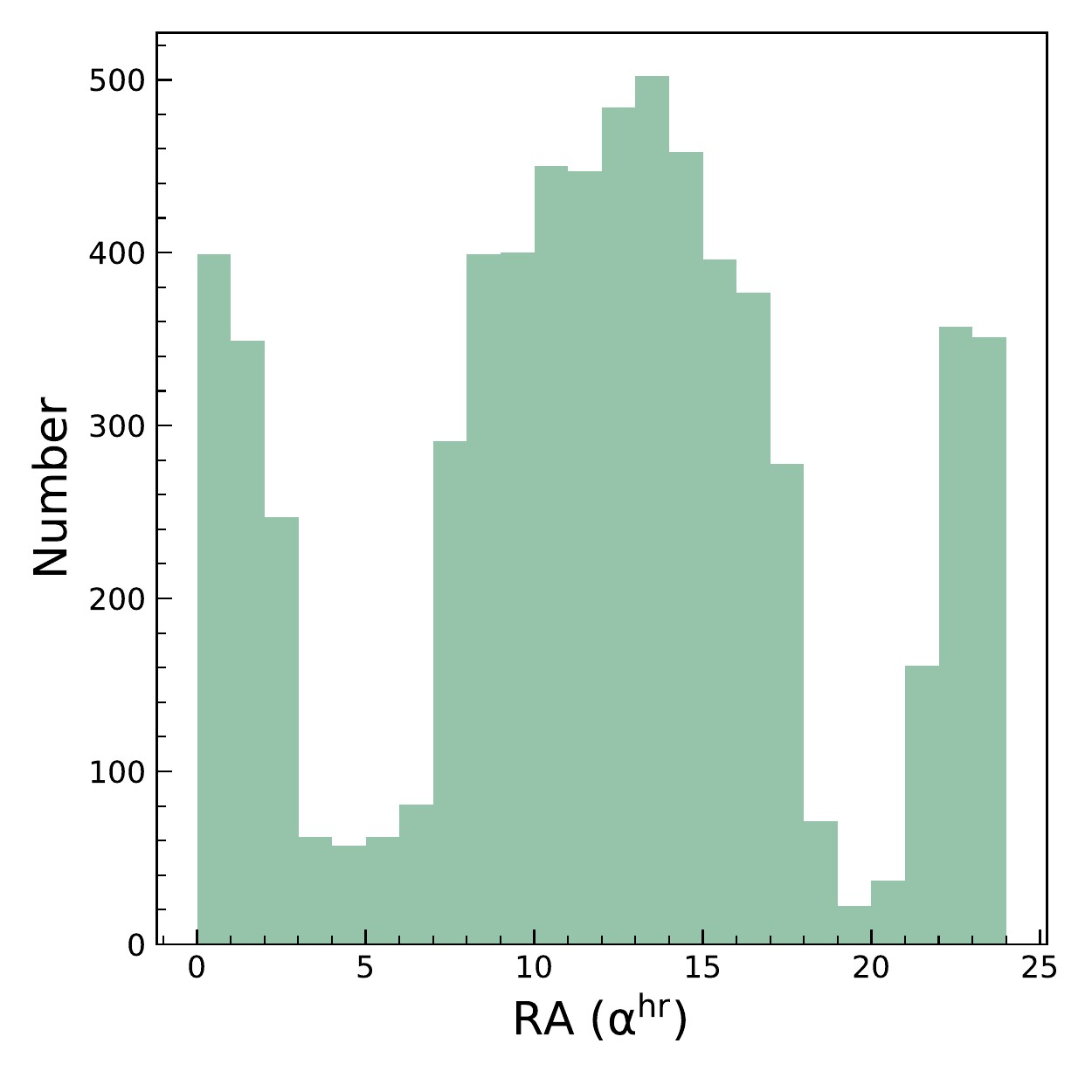}
\caption{Distribution of the selected ILMT quasars in RA.}
	\label{fig:fig-9}
\end{figure}

\subsection{Astrometric calibration of the ILMT field}
The main application of this catalogue of quasars is to calibrate the ILMT 
observations in the world coordinate system. As this catalogue has accurate positions 
from the {\it Gaia}-DR2, with errors in the positions of the order of a few 
$mas$,  using these quasars we expect to achieve sub-arcsec astrometric 
accuracy in the ILMT survey. We performed a Monte Carlo simulation to estimate the astrometric accuracy for the survey. Given the pixel scale of 0.4" and median seeing at Devasthal observatory of 1.1" (Sagar et al. 2000; Sagar, Kumar \& Omar 2019 and references therein), several synthetic CCD frames were generated having a circular Gaussian point spread function at random locations corresponding to different SDSS $i'$ magnitudes assuming a single scan (i.e. exposure time of 102 seconds) as demonstrated by Kumar et al. (2018). The Source Extractor software developed by Barbary (2016) was then used to estimate the centroid of each synthetic source. The $1\sigma$ accuracy in estimating the centroid of a point source having a limiting $i'$ magnitude of 21.4 mag with the 4-m ILMT was found to be 0.09" (see Fig. 8).

The distribution of the ILMT quasars in RA is shown in Fig. 9. This 
Figure indicates that ILMT quasars cover the entire range of RA, however 
for RA between 3$-$6 hr and between 19$-$21 hr the numbers of quasars are around 
60 and 20 per hr angle in the ILMT field, respectively, much lower than the number of quasars in the other 
RA ranges. In this range, we have to compromise with a separate database of astrometric standards such as the Tycho-2 catalogue (Hog et el. 2000) which has an astrometric uncertainty of 0.06" per coordinate and an average star density of $\sim$180 per hr angle. Hence, after adding this uncertainty in quadrature to our estimate the $1\sigma$ positional accuracy for the faintest stars detectable by ILMT will be degraded to 0.11" in the aforementioned RA ranges.

 \subsection{Quasar variability}
 
Optical flux variations in quasars are known since their discovery. They have 
been studied for optical flux variations over a range of time scales 
from minutes to days (Wagner \& Witzel 1995, Ulrich et al. 1997). Most of the 
available studies are limited by the time resolution of the observations 
manifested as gaps in the data. The 4K$\times$4K CCD camera mounted on the ILMT 
can operate over the 4000 to 11000 $\AA$ spectral range in three different SDSS equivalent
filters $g^{\prime}$, $r^{\prime}$ and $i^{\prime}$. The typical exposure time 
for a single frame is $\sim$ 104 s (Surdej et al. 2018). Only one of those filters will be used throughout a single night. The observing strategy of the ILMT will 
enable one to collect good quality data for most of the 6738 quasars that we 
have arrived at in this work primarily for astrometric calibration. Therefore, 
the quasars catalogued in this work can be studied for optical flux variability 
in different optical bands as well as colour variability.  When more epochs of 
observations become available from the ILMT in the future, new candidate quasars 
can also be discovered based on colour-colour diagram (Richards et al. 2002) as well
as optical variability characteristics (Graham et al. 2016).

\subsection{Variability of lensed quasar}
Gravitational lensing, the effect of deflection of light by a foreground intervening compact object (galaxy, cluster, etc.) constitutes a powerful tool that finds applications in many astrophysical areas. Gravitational lensing 
of distant quasars leads to the formation of multiply imaged quasars (Narayan \& Bartelmann 1999; 
Ehlers \& Schneider 1992). For such lensed quasars that show photometric 
variations, it is possible to measure time delays between the lensed quasar images 
by cross-correlating their light curves, 
which in turn can be used to determine the Hubble-Lema\^itre constant $H_{0}$ 
(Refsdal 1964, 1966a), which can help in constraining the dark energy equation of state 
(Kochanek \& Schechter 2004). To date time delays of about 24 lensed quasars are 
known that range from a few days to a few years (Rathna Kumar et al. 2015). Measuring 
such time delays requires long term monitoring of lensed quasars. Among the 
catalogue of quasars arrived at in this work we have only identified one gravitationally lensed 
quasar, namely J1251+295 ($\alpha_{2000}$ = 12:51:07.57, $\delta_{2000}$ = 
29:35:40.50) which has 4 lensed images with maximum angular separation of 
$\sim1.8"$ and can be easily resolved with the ILMT (the median seeing 
at Devasthal site is of the order of 1.1"). The ILMT 
will be able to provide good light curves for this source and many others. Moreover, the ILMT also expects to detect about 50 new multiply imaged quasars 
(Surdej et al. 2018) which opens up the ability of the ILMT to derive more time delays among lensed quasars.

\section{Summary}

This work was aimed to arrive at a catalogue of quasars that could be used as 
calibrators to calibrate the ILMT observations in the world coordinate system. The
details are summarized below.
\begin{enumerate}
\item By cross-correlating the Milliquas catalogue with the {\it Gaia}-DR2, and imposing the condition
of matched sources to have astrometric excess noise significance D $\leq$ 2, we arrived
at a sample of 1047747 quasars over the whole sky. Of these, 6755 quasars are available
in the ILMT stripe.
\item An investigation of the distribution of proper motion for these 6755 quasars have revealed 17 sources ($\sim$0.3\%) to have a PM greater than 20 $mas \, yr^{-1}$.  This confirms that quasars
in the ILMT stripe have PM lesser than 20 $mas \, yr^{-1}$. As the nature
of these 17 objects could not be ascertained due to the lack of optical spectra, they were
excluded from our list.
\item Our final quasar catalogue for the ILMT contains 6738 quasars. Out of which, as per the Milliquas catalogue, 2405 candidates are spectroscopically confirmed type I quasars with broad emission lines, 3 are AGN, 7 are BL Lac objects, 1 is a Type II AGN and 4322 objects are selected through photometric techniques with a probability $>$ 90\% to be quasars. This information has been incorporated in the 8th column of our catalogue (see Table 2). The catalogue that is made
available in this work, in addition to their use as astrometric calibrators, can also 
serve as a large sample for quasar variability studies.
\item We expect to achieve an astrometric accuracy of better than 0.1" in the ILMT survey by using our proposed quasar catalogue.
\end{enumerate}

\section*{Acknowledgements}

We thank the referee for her/his critical review of our manuscript. AKM and RS thank the National Academy of Sciences, India for research grant. AKM specially acknowledges the Universit\'e de Li\`ege for providing the scholarship "Erasmus + International Credit Mobility".  AKM, CSS and RS acknowledge the Alexander von Humboldt Foundation, Germany for the award of Group linkage long-term research program between IIA, Bengaluru and European Southern Observatory, Garching, Germany. AKM and RS are thankful to the Director, IIA for providing institutional infrastructural support during this work. This research has used data from the SDSS and {\it Gaia}, operated by the European Space Agency.


\clearpage


\vspace{-1em}

\begin{theunbibliography}{} 
\vspace{-1.5em}





\bibitem{latexcompanion}
Abolfathi B., et al., 2018, ApJS, 235, 42

\bibitem{latexcompanion}
Bovy J., et al., 2011, The Astrophysical Journal, 729, 141

\bibitem{latexcompanion}
Barbary, K. (2016). SEP: Source Extractor as a library. Journal of Open Source Software, 1(6), 58.

\bibitem{latexcompanion}
Croom S., et al., 2004, in M\'ujica R., Maiolino R., eds, Multiwavelength AGN Surveys. pp 57–62, doi:10.1142/9789812702432 0015


\bibitem{latexcompanion}
Ehlers J., Schneider P., 1992, Gravitational Lensing. p. 1,
doi:10.1007/3-540-56180-3 1

\bibitem{latexcompanion}
Flesch E. W., 2015, Publ. Astron. Soc. Australia, 32, e010

\bibitem{latexcompanion}
Flesch E. W., 2017, VizieR Online Data Catalog, p. VII/280

\bibitem{latexcompanion}
Gaia Collaboration Mignard F., Klioner S., Lindegren L., Hernandez J.,
Bastian U., Bombrun A., 2018, Astronomy and Astrophysics, 616, A14

\bibitem{latexcompanion}
Graham M., Djorgovski S. G., Stern D., Drake A. J., Mahabal
A. A., Glikman E., 2016, in American Astronomical Society

\bibitem{latexcompanion}
Hog, E., et al. The Tycho-2 catalogue of the 2.5 million brightest stars. NAVAL OBSERVATORY WASHINGTON DC, 2000.

\bibitem{latexcompanion}
Kochanek C. S., Schechter P. L., 2004, in Freedman W. L., ed., Measuring and Modeling the Universe. p. 117 (arXiv:astroph/0306040)

\bibitem{latexcompanion}
Kumar, B., Pandey, K. L., Pandey, S. B., Hickson, P., Borra, E. F., Anupama, G. C., \& Surdej, J. (2018). The zenithal 4-m International Liquid Mirror Telescope: a unique facility for supernova studies. Monthly Notices of the Royal Astronomical Society, 476(2), 2075-2085.

\bibitem{latexcompanion}
Lindegren L., et al., 2018, Astronomy and Astrophysics, 616, A2

\bibitem{latexcompanion}
Lindegren L., Lammers U., Hobbs D., O’Mullane W., Bastian U.,
Hern\'andez J., 2012, A\&A, 538, A78

\bibitem{latexcompanion}
Marrese P. M., Marinoni S., Fabrizio M., Altavilla G., 2019, A\&A,
621, A144

\bibitem{latexcompanion}
Myers A. D., et al., 2015, ApJS, 221, 27


\bibitem{latexcompanion}
Narayan R., Bartelmann M., 1999, in Dekel A., Ostriker J. P.,
eds, Formation of Structure in the Universe. p. 360


\bibitem{latexcompanion} 
P\^aris I., et al., 2018, A\&A, 613, A51

\bibitem{latexcompanion}
Peters C. M., et al., 2015, ApJ, 811, 95



\bibitem{latexcompanion}
Rathna Kumar S., Stalin C. S., Prabhu T. P., 2015, A\&A, 580,
A38

\bibitem{latexcompanion}
Refsdal S. 1964, MNRAS, 128, 4, p. 307-310

\bibitem{latexcompanion}
Refsdal, S. 1966a, MNRAS, 132, 101


\bibitem{latexcompanion}
Richards G. T., et al. 2002, AJ, 123, 2945


\bibitem{latexcompanion}
Richards G. T., et al., 2009, ApJS, 180, 67



\bibitem{latexcompanion}
Richards G. T., et al., 2015, VizieR Online Data Catalog, p.
J/ApJS/219/39


\bibitem{latexcompanion}
Ross A. J., et al., 2011, MNRAS, 417, 1350



\bibitem{latexcompanion}
Sagar et al. 2000, Astron. Astrophysics Suppl. Vol 114, p. 349-362.

\bibitem{latexcompanion}
Sagar, R. Kumar, B. \& Omar, A. 2019 Current Science Vol. 117, p. 365

\bibitem{latexcompanion}
Schmidt M., 1963, Nature, 197, 1040

\bibitem{latexcompanion}
Souchay J., et al., 2015, A\&A, 583, A75

\bibitem{latexcompanion}
Surdej J., et al., 2018, Bulletin de la Societe Royale des Sciences
de Liege, 87, 68

\bibitem{latexcompanion}
Ulrich M., Maraschi L., Urry C.M., 1997, ARA\&A, 35, 445


\bibitem{latexcompanion}
Varshni Y. P., 1982, SScT, 521, 532

\bibitem{latexcompanion}
V\'eron-Cetty M. P., V\'eron P., 2006, A\&A, 455, 773


\bibitem{latexcompanion} 
V\'eron-Cetty M.-P., V\'eron P., 2010, A\&A, 518, A10 



\bibitem{latexcompanion} 
Wagner S.J., Witzel A., 1995, ARA\&A, 33, 163 


\end{theunbibliography}
\end{document}